\documentclass[final,3p,times,twocolumn]{elsarticle}

\usepackage{amsfonts}
\usepackage{amsmath}
\usepackage{graphicx}
\usepackage{bm}
\usepackage{amssymb}
\usepackage{dcolumn}
\usepackage{color,ulem}
\usepackage{epstopdf}

\begin{document}

\title{What determines the sign of the spin Hall effects in Cu alloys doped with 5d elements?}

\author{Zhuo Xu$^{1}$,
Bo Gu$^{1}$, Michiyasu Mori$^{1}$, Timothy Ziman$^{2,3}$, and Sadamichi Maekawa$^{1,4}$}

\address{$^{1}$Advanced Science Research Center, Japan Atomic Energy Agency, Tokai 319-1195, Japan
\\ $^{2}$Institut Laue Langevin, 71 avenue des Martys, CS-20156, F-38042 Grenoble Cedex 9, France
\\ $^{3}$LPMMC (UMR 5493),CNRS and  Universit\'{e} Joseph-Fourier Grenoble, 38042 Grenoble, France
\\ $^{4}$ERATO, Japan Science and Technology Agency, Sendai 980-8577, Japan
}

\date{\today}
\begin{abstract}

We perform a systematical analysis of the spin Hall effect (SHE) in the Cu alloys doped with a series of $5d$ elements, by the combined approach of density functional theory and Hartree-Fock approximation. We find that not only the spin orbit interactions (SOI) in both the $5d$ and $6p$ orbitals, but also the local correlations in the $5d$ orbitals of the impurities, are decisive on the sign of the spin Hall angle (SHA). Including all of these three factors properly, we predict the SHA for each alloy in the series. The signs of CuIr and CuPt are sensitive to perturbation of the local correlations. This observation is favorable for controlling the sign of the transverse spin Hall voltage.

\end{abstract}


\maketitle

\section{Introduction}

Spintronics is a rich research field, not only for the wide applications of low energy
consumption and energy transformation devices, but also for the new physics with the
interplays among charge, spin, orbital, heat and so on \cite{Maekawa2}.
The spin Hall effect (SHE), which converts the injected longitudinal charge current into
the transverse spin current via the spin-orbit interaction (SOI),
is crucial for the development of spintronic devices.
The SHE is characterized by the ratio between the transverse and the longitudinal resistivities
(or conductivities), called spin Hall angle (SHA).
The magnitude of SHA describes conversion efficiency between the charge current and the spin current,
while the sign distinguishes the scattering direction of electrons, i.e.,
clockwise or anticlockwise into the transverse direction.
It is well known that to raise the magnitude of SHA is the key to enhance the
whole efficiency of the devices based on the SHE \cite{Saitoh1,Saitoh2}.
However, the sign of SHA is not yet utilized as a degree of freedom in spintronic devices.

In the experiment of the dilute CuIr alloys,
the dominant contribution to the SHE was verified to be an extrinsic skew scattering mechanism,
and the SHA was measured to be positive 2.1\% \cite{Niimi}.
We have found out in theory that including the local correlation effects of the $5d$ orbitals of Ir is
decisive to obtain a positive sign of SHA consistent with experiment,
and a small change of the $5d$ electron number of Ir
with the local correlations may change the sign \cite{Xu1,Xu2}.

On the other hand,
among the previous theoretical works about the SHE of the Cu alloys doped with $5d$ elements,
Fert and Levy estimated the SHA by the atomic electron numbers of $5d$ orbitals with SOI,
obtaining a sign change of SHA in the middle of the series of $5d$ impurities \cite{FertLevy}.
In contrary, based on the $ab$ $initio$ calculations, including the SOI in both $5d$ and $6p$ orbitals,
Fedorov et al. obtained an uniform sign of SHA contributed by the skew scattering
among the series of $5d$ impurities from Lu to Pt \cite{Fedorov}.

In the present work, we analyze the SHE among the dilute Cu alloys with a series of $5d$ elements as impurities,
and find the key factors which are decisive on the sign of SHA.
Furthermore, we screen out the alloys whose signs of SHA are sensitive to perturbation of the local correlations as CuIr,
which will be favorable for the sign control of SHE.

\section{Theoretical Approach}

The extrinsic SHE is caused by the spin-orbit interactions (SOI) in the orbitals of the impurities.
As one of the mechanisms of extrinsic SHE,
the skew scattering generates the spin Hall resistivity linear with the impurity concentration in the dilute alloys.
Based on the Anderson model \cite{Anderson},
the contribution of skew scattering on the extrinsic SHE in the nonmagnetic Cu alloys
with dilute $5d$ impurities can be described by
a single-impurity multi-orbital model, including
the SOI in both the $6p$ orbitals $\zeta$ and the $5d$ orbitals $\xi$ of the impurity
with the parameters $\lambda_{p}$ and $\lambda_{d}$, respectively,
together with the local correlations of
the on-site Coulomb repulsion $U$ ($U^{\prime}$) within (between) the $5d$ orbitals,
and the Hund coupling $J$ between the $5d$ orbitals of the impurity \cite{Xu2}:
\begin{equation}
\begin{split}
H_{0}=&\sum_{\textbf{k},\alpha,\sigma}\epsilon_{\alpha\textbf{k}}
  c^{\dag}_{\textbf{k}\alpha\sigma}c_{\textbf{k}\alpha\sigma},\\
&+\sum_{\textbf{k},\alpha,\beta,\sigma}(V_{\beta\textbf{k}\alpha }
    d^{\dag}_{\beta\sigma} c_{\textbf{k}\alpha\sigma} + \textrm{H.c.})
   + \sum_{\beta,\sigma}\epsilon_{\beta}d^{\dag}_{\beta\sigma}d_{\beta\sigma},\\
&+\frac{\lambda_{p}}{2}\sum_{\zeta\sigma,\zeta^{\prime}\sigma^{\prime}}d^{\dagger}_{\zeta\sigma}
   (\textbf{l})_{\zeta\zeta^{\prime}}\cdot(\pmb\sigma)_{\sigma\sigma^{\prime}}d_{\zeta^{\prime}\sigma^{\prime}}\\
   &+\frac{\lambda_{d}}{2}\sum_{\xi\sigma,\xi^{\prime}\sigma^{\prime}}d^{\dagger}_{\xi\sigma}
   (\textbf{l})_{\xi\xi^{\prime}}\cdot(\pmb\sigma)_{\sigma\sigma^{\prime}}d_{\xi^{\prime}\sigma^{\prime}},
\end{split}
\label{andersonmodel0}
\end{equation}
\begin{equation}
\begin{split}
H=&H_{0}+U\sum_{\xi}n_{\xi\uparrow}n_{\xi\downarrow}\\
  & + \frac{U^{\prime}}{2}\sum_{\xi\neq\xi',\sigma,\sigma^{\prime}}
     n_{\xi\sigma}n_{\xi'\sigma^{\prime}}
   - \frac{J}{2}\sum_{\xi\neq\xi',\sigma}n_{\xi\sigma}n_{\xi'\sigma}.
\end{split}
\label{andersonmodel}
\end{equation}
In Eq.(\ref{andersonmodel0}), $\epsilon_{\alpha\textbf{k}}$ is the energy band $\alpha$ of the Cu host,
$\epsilon_{\beta}$ is the energy level of the orbital $\beta$ of the $5d$ impurity,
and $V_{\beta,\alpha}(\textbf{k})$ is the hybridization between the orbital $\beta$
of the impurity and the band $\alpha$ of the host.
In Eq.(\ref{andersonmodel}),
the local correlations are included only within the localized $5d$ orbitals $\xi$ of the impurity,
employing the relations of $U=U^{\prime}+2J$ \cite{Maekawa1} and $J/U=0.3$.

In the orbital with the orbital angular momentum $l$,
the SOI split the states of the orbital into two groups of degenerated states
with the total angular momentum $j=l\pm\frac{1}{2}$ and the degeneracy $D_{l\pm}=2j+1$.
In the Cu alloys with the $5d$ impurities including SOI,
according to the the Friedel sum rule \cite{Fert,Langreth},
the phase shifts $\delta_{l}^{\pm}$ can be calculated
from the occupation numbers $N_{l\pm}$ of the spin-orbit split states of the impurity and the Cu host,
\begin{equation}
\delta_{l}^{\pm} = \pi(N_{l\pm}^{imp}-N_{l\pm}^{Cu})/D_{l\pm},
\label{phaseshifts}
\end{equation}
where $l$=1 corresponds to the $6p$ states of the impurity and the $4p$ states of Cu,
$l$=2 corresponds to $5d$ states of the impurity and the $3d$ states of Cu, respectively.
The total occupation number of the $5d$ states of the impurity $N_{d}^{imp}$ can be obtained by
\begin{equation}
N_{d}^{imp}=N_{2+}^{imp}+N_{2-}^{imp}.
\label{ntotal}
\end{equation}
The occupation number of each of the degenerate $5d$ states of the impurity $n_{d\pm}$ can be obtained by
\begin{equation}
n_{d\pm}=N_{2\pm}^{imp}/D_{2\pm},
\label{neach}
\end{equation}
where $n_{d\pm}$ will be between 0 and 1.
The occupation numbers of the impurities are defined via
projections of the occupied states onto the Wannier states centered at the impurities as point defects
and extended in the whole supercell.
In the dilute alloys, a point defect cannot be charged in metal.
Thus the total occupation numbers of the $6s$, $6p$ and $5d$ states of the impurities
are conserved to the atomic valence electron numbers \cite{Xu1,Xu2},
\begin{equation}
N_{s}^{imp}+N_{p}^{imp}+N_{d}^{imp} = C^{imp},
\label{charge}
\end{equation}
where the constant $C^{imp}$=4, 5, 6, 7, 8, 9, 10, 11 and 12,
for the impurities of Hf, Ta, W, Re, Os, Ir, Pt, Au and Hg, respectively.

Based on the model in Eq.(\ref{andersonmodel0}),
by the Hartree-Fock (HF) approximation,
the $5d$ states of the impurities can be considered as virtual bound states with a width parameter $\Delta$,
and the $5d\pm$ states correspond to the energy levels of $\epsilon_{0,d\pm}$, respectively.
It has the relations of \cite{Xu2,Anderson}
\begin{equation}
\Delta \cot(\pi n_{d\pm})=\epsilon_{0,d\pm}-\epsilon_{F}=E_{0,d\pm},
\label{sc1}
\end{equation}
where the $\epsilon_{F}$ is the Fermi level and
$E_{0,d\pm}$ are the energy levels of the $5d\pm$ states relative to the Fermi level.
While the local correlations are included in the $5d$ states of the impurity as in Eq.(\ref{andersonmodel}),
it has the self-consistent relations of \cite{Xu2}
\begin{equation}
\begin{split}
E_{d\pm}=&\Delta \cot(\pi n_{d\pm})\\
=&E_{0,d\pm}+U(\frac{3}{5}n_{d+}+\frac{2}{5}n_{d-})+U^{\prime}(\frac{24}{5}n_{d+}+\frac{16}{5}n_{d-})\\
&-J(\frac{12}{5}n_{d+}+\frac{8}{5}n_{d-}),
\end{split}
\label{sc2}
\end{equation}
which will raise the $5d$ energy level and lower the $5d$ occupation numbers of the impurity \cite{Xu1,Xu2,Anderson}.
The decrease of $N_{d}^{imp}$ will induce the increase of $N_{p}^{imp}$ according to Eq.(\ref{charge}),
which can be approximately estimated by fixing the ratios of $N_{p}^{imp}/N_{s}^{imp}$ and $N_{1+}^{imp}/N_{1-}^{imp}$.
The corresponding phase shifts can be obtained by Eq.(\ref{phaseshifts}).

For the SHE contributed by the skew scattering in Cu alloys with $5d$ impurities,
the SHA defined in terms of resistivity $\rho$ \cite{Xu1,GuCuBi} can be calculated
from the phase shifts $\delta_{1}^{\pm}$ of the $p\pm$ channels and
$\delta_{2}^{\pm}$ of the $d\pm$ channels.
Including the SOI in both the $p$ and $d$ channels, the SHA $\Theta$ is obtained by \cite{Xu2}
\begin{equation}
\begin{split}
\Theta & (\delta_{1}^{+},\delta_{1}^{-},\delta_{2}^{+},\delta_{2}^{-}) = A/B,\\
A=& -2[9\sin(\delta^{+}_{1}-\delta^{+}_{2})\sin\delta^{+}_{1}\sin\delta^{+}_{2}\\
&-4\sin(\delta^{+}_{1}-\delta^{-}_{2})\sin\delta^{+}_{1}\sin\delta^{-}_{2}\\
&-5\sin(\delta^{-}_{1}-\delta^{-}_{2})\sin\delta^{-}_{1}\sin\delta^{-}_{2}],\\
B=& 45\sin^2\delta^{+}_{2}+30\sin^2\delta^{-}_{2}+50\sin^2\delta^{+}_{1}+25\sin^2\delta^{-}_{1}\\
&+6\sin\delta^{+}_{1}\sin(2\delta^{+}_{2}-\delta^{+}_{1})+12\sin\delta^{-}_{1}\sin(2\delta^{+}_{2}-\delta^{-}_{1})\\
&+14\sin\delta^{+}_{1}\sin(2\delta^{-}_{2}-\delta^{+}_{1})-2\sin\delta^{-}_{1}\sin(2\delta^{-}_{2}-\delta^{-}_{1}).
\end{split}
\label{SHA0}
\end{equation}
When the SOI in the $p$ (or $d$) channels are omitted,
with $\delta_{1}=\delta_{1}^{+}=\delta_{1}^{-}$ (or $\delta_{2}=\delta_{2}^{+}=\delta_{2}^{-}$),
the simplified formula for SHA $\Theta_{d}(\delta_{1},\delta_{1},\delta_{2}^{+},\delta_{2}^{-})$
(or $\Theta_{p}(\delta_{1}^{+},\delta_{1}^{-},\delta_{2},\delta_{2})$) can be obtained from Eq.(\ref{SHA0}).

Based on the model in Eq.(\ref{andersonmodel}) and the formula in Eq.(\ref{SHA0}),
to calculate the SHA for the Cu alloys with a series of $5d$ impurities,
we employ the combined method of the density functional theory (DFT) and
HF approximation, as we have done before for the CuIr alloys \cite{Xu2}.

\section{Results and Discussions}

\begin{figure}[tbp]
\centering
\includegraphics[width = 7 cm]{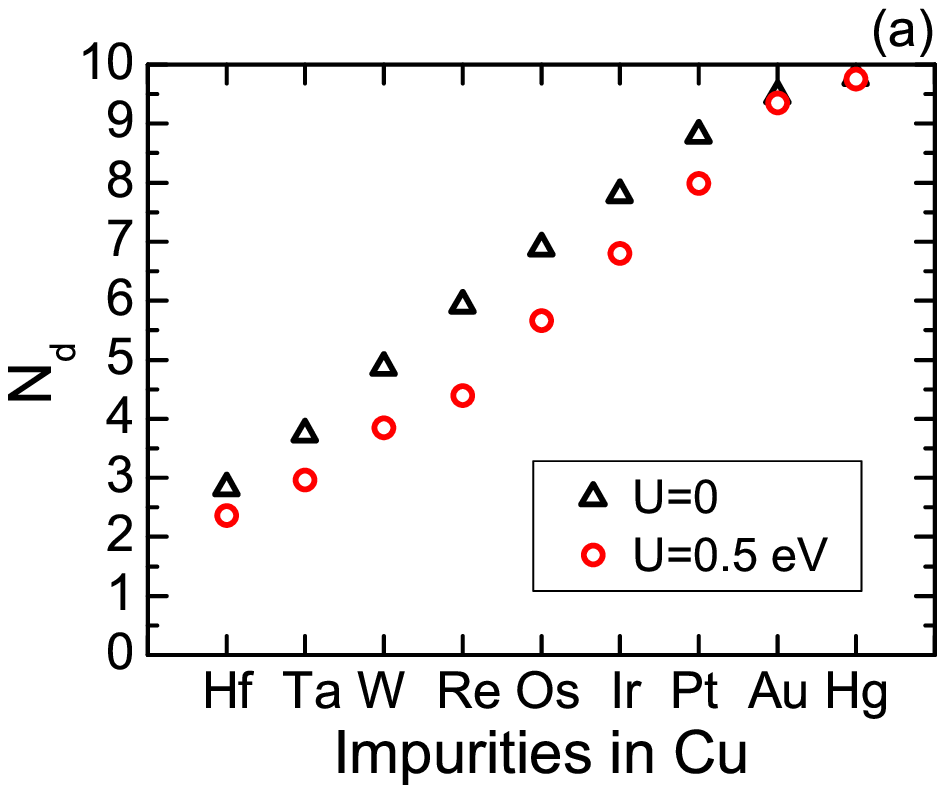}
\includegraphics[width = 7 cm]{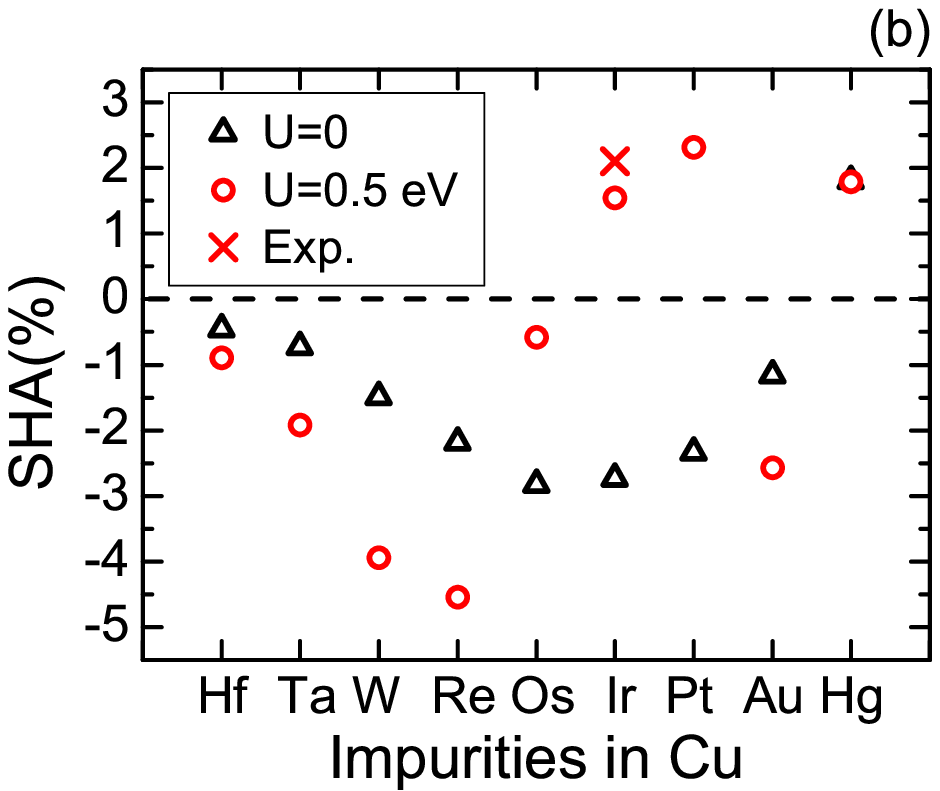}
\caption{(a) The occupation numbers of the $5d$ orbitals of the impurities $N_{d}$ and
(b) the SHA ($\Theta(\delta_{1}^{+},\delta_{1}^{-},\delta_{2}^{+},\delta_{2}^{-})$)
of the Cu alloys with $5d$ impurities
under the local correlation U=0 (black triangles) and U=0.5 eV (red circles) in $5d$ orbitals, respectively.
The experimental data of SHA in CuIr \cite{Niimi} is marked by the red cross in (b).}
\label{hf1}
\end{figure}

\begin{figure}[tbp]
\centering
\includegraphics[width = 7 cm]{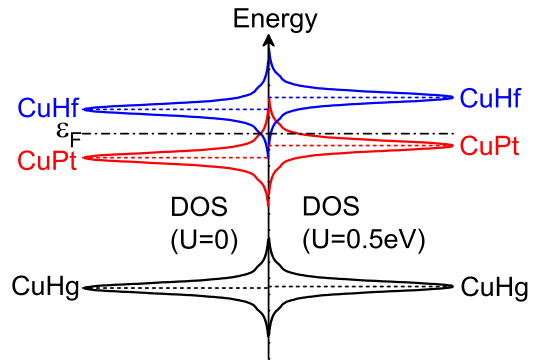}
\caption{Schematic picture of the total density of states (DOS) of the $5d$ virtual bound states,
for the impurities in the dilute alloys CuHf (blue), CuPt (red) and CuHg (black),
with the on-site Coulomb repulsion $U$=0 and $U$=0.5 eV, respectively.
$\varepsilon_{F}$ is the Fermi level.
The amount of the shift by $U$ depends on the atomic number of the impurities.}
\label{anderson5d}
\end{figure}

For the dilute alloys of Cu with a series of $5d$ impurities,
including the SOI in both the $5d$ and $6p$ orbitals,
the $5d$ occupation numbers $N_{d}$ and the SHA
are calculated by the DFT+HF method following Eqs.(\ref{sc2}) and (\ref{SHA0}),
without the local correlations ($U$=0) or with the local correlations of $U$=0.5 eV in the $5d$ orbitals of the impurities,
respectively, as shown in Fig.{\ref{hf1}}.
Estimates from experiments give the parameter of the on-site Coulomb repulsion $U$
of $5d$ orbitals in the range of 0-1.0 eV for pure metals \cite{UPt},
0-1.5 eV for insulating compounds \cite{Savrasov},
less than 0.4 eV for the ferromagnetic metal of osmates \cite{Savrasov},
about 0.5 eV for iridates \cite{UIr},
and 0.5-0.6 eV for Pt alloys \cite{UPt}.
Below we take $U$=0.5 eV to compare the whole series of $5d$ impurities in Cu alloys.

As $U$ switches from 0 to 0.5 eV,
according to Eq.(\ref{andersonmodel}) based on the Anderson model \cite{Xu1,Xu2,Anderson},
the $5d$ state will be shifted up, accompanied by a decrease of $N_{d}$.
The amount of the shift varies among the $5d$ impurities,
as intuitively shown in Fig.{\ref{anderson5d}}.
The $N_{d}$ lowered by $U$ for each of the $5d$ impurities are marked in Fig.{\ref{hf1}}(a).
In the middle of the series with $N_{d}$ around the half-filling value of 5,
the $N_{d}$ is obviously affected by $U$,
corresponding to the schematic picture in Fig.{\ref{anderson5d}}
that the peaks of the $5d$ band from Hf to Pt are relatively close to the Fermi level.

\begin{figure*}[t]
\centering
\includegraphics[width = 0.8\linewidth]{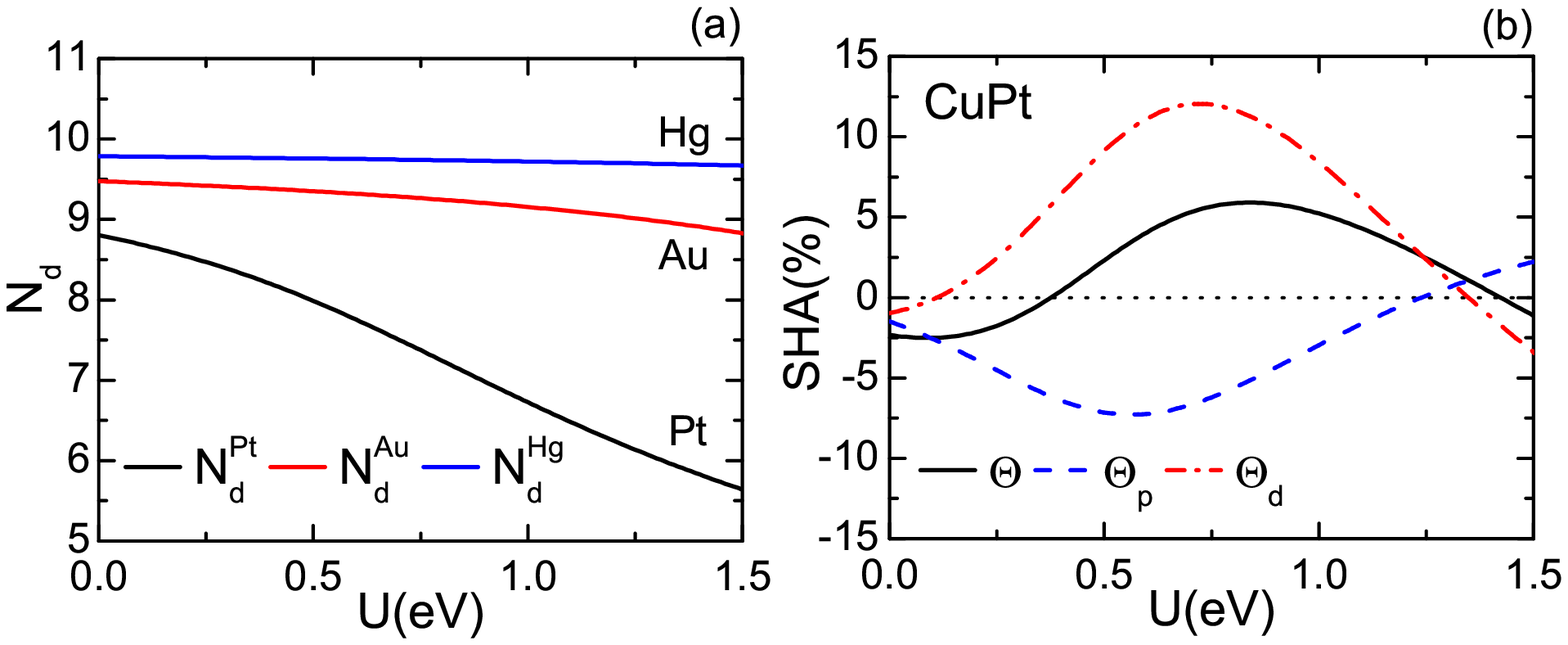}
\includegraphics[width = 0.8\linewidth]{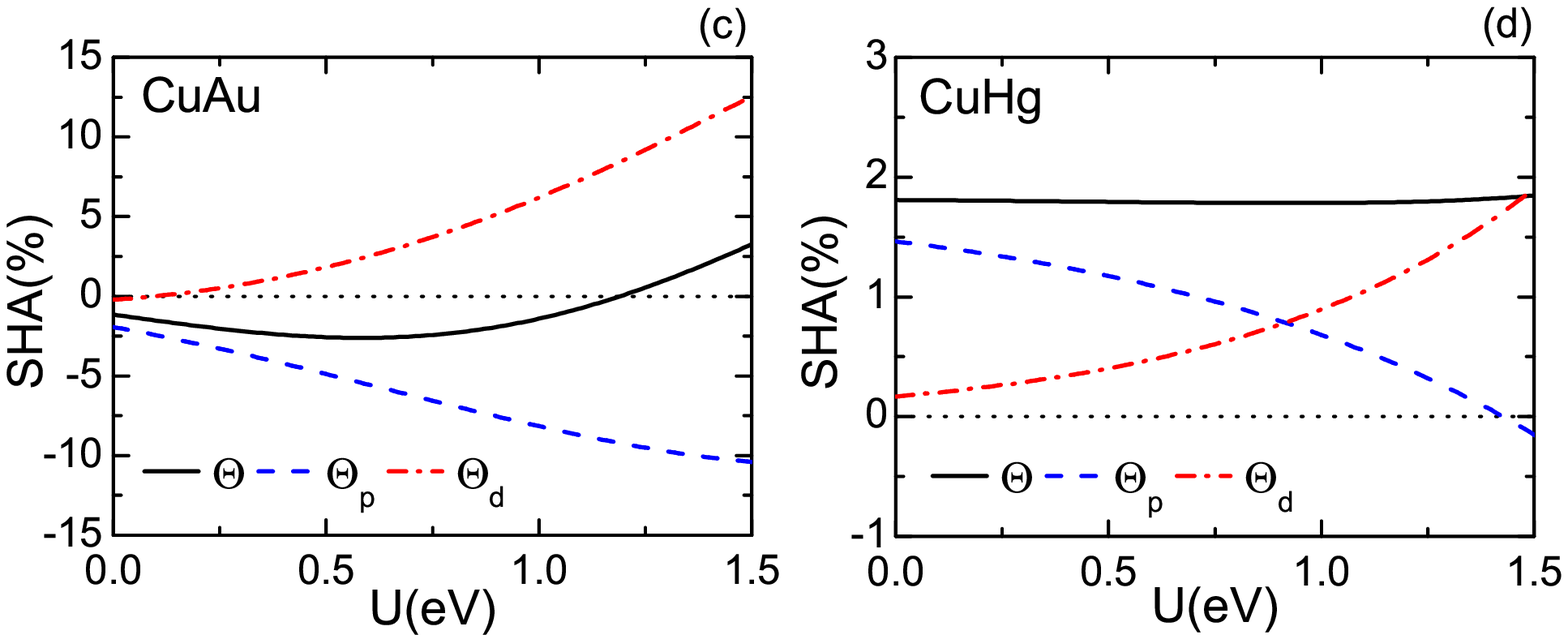}
\caption{(a) The occupation numbers of the $5d$ orbitals $N_{d}$ of
Pt (black curve), Au (red curve) and Hg (blue curve) in the Cu as functions of the local correlation $U$ in $5d$ orbitals.
The SHA of the (b) CuPt, (c) CuAu and (d) CuHg alloys as functions of $U$.
In (b), (c) and (d), the functions of SHA $\Theta(\delta_{1}^{+},\delta_{1}^{-},\delta_{2}^{+},\delta_{2}^{-})$,
$\Theta_{p}(\delta_{1}^{+},\delta_{1}^{-},\delta_{2},\delta_{2})$ and $\Theta_{d}(\delta_{1},\delta_{1},\delta_{2}^{+},\delta_{2}^{-}$)
are denoted by black solid, blue dash and red dash-dot curves, respectively.}
\label{hf2}
\end{figure*}

Previous theories leave the following discrepancy regarding the sign of the SHA
in the Cu alloys with $5d$ impurities contributed by the skew scattering.
Fert and Levy gave a sign change of SHA in the middle of the $5d$ series,
including the SOI only in the $5d$ orbitals without local correlations,
and taking the phase shift of $p$ channel as a parameter \cite{FertLevy}.
Fedorov et al. obtained that the signs of SHA among a series of $5d$ impurities from Hf to Pt do not change,
including the SOI in both the $5d$ and $6p$ orbitals but without local correlations.
Their result for CuIr is opposite in sign to the experiment under the consistent definition of SHA via resistivity \cite{Fedorov}.

In comparison, in Fig.{\ref{hf1}}(b),
the series of SHA under $U$=0 is very close to
the $ab$ $initio$ results of the Cu alloys with the impurities from Ta to Pt in Ref.\cite{Fedorov}
(with the SHA consistently defined via resistivity), as expected.
However, the series of SHA deviate obviously between $U$=0 and $U$=0.5 eV,
except for the CuHg case with the $5d$ states nearly fully occupied.
It is remarkable that for the cases of CuIr and CuPt,
the SHA changes sign between $U$=0 and $U$=0.5 eV.
Moreover, with $U$=0.5 eV, CuIr gives a SHA of +1.6\%,
consistent to experimental value of +2.1\% \cite{Niimi}, as marked in Fig.{\ref{hf1}}(b).
These facts demonstrate that the SOI in both the $5d$ and $6p$,
together with the local correlations in the $5d$ orbitals of the impurities,
are decisive on the sign of SHA of the Cu alloys.
All of these three factors are necessary to be included properly to estimate the SHA contributed by the skew scattering.

Fig.{\ref{hf2}} shows the changes of $N_{d}$ and SHA along with $U$ for the cases of dilute CuPt, CuAu and CuHg alloys.
The results of CuPt in Figs.{\ref{hf2}}(a) and (b) are quite similar to the results of CuIr shown in Ref.\cite{Xu2}.
Close to the generally realistic value of $U$=0.5 eV,
both the CuIr \cite{Xu2} and CuPt have a sign change of SHA between $U$=0.3 and 0.4 eV,
corresponding to the change of $N_{d}$ from $U$=0.5 eV by as few as about 0.3 electrons.
In the experiment of the heavily doped CuPt alloys with 10\% of Pt,
the charge transfer into the $5d$ orbitals of Pt is 0.25 electrons \cite{Oh}.
If it were possible to devise a way to change the $N_{d}$ in CuIr or CuPt by about 0.3 electrons,
or if the $U$ could be altered by about 0.1 eV,
the sign of SHE could be externally controlled.

Within the range of $U$ from 0 to 1.5 eV,
the sign of SHA of the Cu alloys with $5d$ impurities from Hf to Os is stably negative.
There are more sign changes of SHA around $U$=1.25 eV for CuPt and CuAu,
as shown in Figs.{\ref{hf2}}(b) and (c),
but such a large $U$ is unlikely to be relevant for $5d$ impurities.
In contrast to CuPt, the changes of $N_{d}$ and SHA in CuAu and CuHg are extremely slow with $U$,
as shown in Figs.{\ref{hf2}}(a), (c) and (d).
This is because the $5d$ states of Au and Hg are nearly fully occupied,
and the center of $5d$ band is so deep below the Fermi level that
the shift by $U$ induces little change of $N_{d}$, as shown in Fig.{\ref{anderson5d}}.
Increasing $U$ from 0 to 0.5 eV,
the decrease of $N_{d}$ of Au and Hg is as small as 0.13 and 0.03 electrons, respectively.
Also due to the nearly full occupation of the $5d$ states,
the phase shifts of $d$ channel are very close to zero
and the spin-orbit splits between $d+$ and $d-$ states are very small for CuAu and CuHg.
As shown in Figs.{\ref{hf2}}(c) and (d), with the $U$ from zero to around 0.5 eV,
the SHA function $\Theta(\delta_{1}^{+},\delta_{1}^{-},\delta_{2}^{+},\delta_{2}^{-})$
is closer to $\Theta_{p}(\delta_{1}^{+},\delta_{1}^{-},\delta_{2},\delta_{2})$,
rather than $\Theta_{d}(\delta_{1},\delta_{1},\delta_{2}^{+},\delta_{2}^{-}$),
indicating that the skew scattering by the SOI in $6p$ orbitals overwhelms the contribution by the SOI in $5d$ orbitals.

In summary, by the combined method of DFT+HF, we analyze the SHE in the Cu alloys with a series of $5d$ impurities.
We stress that it is necessary to include the SOI in both the $6p$ and $5d$ orbitals,
together with the local correlations in the $5d$ orbitals of the impurity,
to estimate the SHA of these alloys.
The results reveal that the sign of SHA of CuIr and CuPt would be sensitive to perturbation of the local correlations,
which is favorable for the sign control of SHE performing as a spin current switch.

\section*{Acknowledgement}	
This work is supported by Grant-in-Aid for Scientific Research (Grant No.25287094, No.26108716, No.26247063, No.26103006, No.15K05192, and No.15K03553) from JSPS and MEXT, by the National Science Foundation under Grant No. NSF PHY11-25915, by the REIMEI project of JAEA, and by the inter-university cooperative research program of IMR Tohoku University.


\end{document}